\documentclass[conference]{IEEEtran}
\IEEEoverridecommandlockouts
\usepackage{cite}
\usepackage{amsmath,amssymb,amsfonts}
\usepackage{algorithmic}
\usepackage{graphicx}
\usepackage{textcomp}
\usepackage{xcolor}
\def\BibTeX{{\rm B\kern-.05em{\sc i\kern-.025em b}\kern-.08em
    T\kern-.1667em\lower.7ex\hbox{E}\kern-.125emX}}

\usepackage{multirow}
\usepackage{makecell}

\begin{document}
\title{Systematic Study of Dysarthric Speech Recognition: Spectral Features and Acoustic Models}


\author{
    \IEEEauthorblockN{Paban Sapkota\IEEEauthorrefmark{1}, Hemant Kumar Kathania\IEEEauthorrefmark{1},
    Mikko Kurimo\IEEEauthorrefmark{2},
    Sudarsana Reddy Kadiri\IEEEauthorrefmark{3}, and Shrikanth Narayanan\IEEEauthorrefmark{3}}
    \\
    \IEEEauthorblockA{\IEEEauthorrefmark{1} Department of Electronics and Communication Engineering, National Institute of Technology Sikkim, India.  \\
    Emails: phec230006@nitsikkim.ac.in, hemant.ece@nitsikkim.ac.in}

    \IEEEauthorblockA{\IEEEauthorrefmark{2} Department of Information and Communications Engineering, Aalto University, Finland.\\
Email: mikko.kurimo@aalto.fi}
    \IEEEauthorblockA{\IEEEauthorrefmark{3} Signal Analysis and Interpretation Laboratory (SAIL), University of Southern California, Los Angeles, USA.\\
    Emails: skadiri@usc.edu, shri@usc.edu}
}

\maketitle

\begin{abstract}

The challenge associated with recognizing dysarthric speech primarily arises from pronounced acoustic variability attributed to impaired articulatory precision. Past research has demonstrated improved recognition through the use of hybrid DNN/HMM sequence discriminative training. This paper presents a comprehensive investigation of various combinations of acoustic features tailored to different Acoustic Models, offering suitable feature selections for each. The incorporation of Pitch features notably improved recognition performance, especially for sentence recognition tasks involving dysarthric speech. Through a systematic examination of the TORGO database, we have demonstrated the potential to enhance the performance of the state-of-the-art Factorized Time Delay Neural Network (F-TDNN) model for recognizing dysarthric speech. Our methods, implemented with the F-TDNN model, resulted in a 4.65\% relative improvement in isolated word recognition and a 4.63\% relative improvement in sentence recognition for dysarthric speech, compared to previous research. This improvement effectively compensates for speech variability, attributable to our deliberate selection of the number of overlapping frames between consecutive training example chunks.

\end{abstract}

%
\begin{IEEEkeywords}
Dysarthria, speech variability, isolated word recognition, sentence recognition, F-TDNN, overlapping frames.
\end{IEEEkeywords}
\vspace{-0.3cm}
\section{Introduction}\vspace{-0.1cm}
\label{sec:intro}
Dysarthria is a speech disorder characterized by difficulties in controlling the neuro-motor functions responsible for speech articulation \cite{dysar}. Automatic Speech Recognition (ASR) systems are undergoing rapid development and are progressing toward the point where they have the potential to achieve human-level speech perception capabilities. It is essential to note that this progress primarily pertains to ASR systems designed for typical speech from control or healthy speakers. While a considerable amount of effort has been dedicated to the field of Dysarthric Automatic Speech Recognition (DyASR) \cite{review}, there remains a significant need to reach an acceptable level of ASR performance.

Dysarthric speech differs from typical speech in terms of acoustic properties and pronunciation distinctions. Challenges in accurately articulating sounds and forming words result in a notable lack of clarity and comprehensibility, leading to poor intelligibility. In DyASR, there are significant intra-speaker disparities and even greater inter-speaker disparities, constituting two primary challenges. Collecting dysarthric speech data introduces another major dimension of challenge, primarily due to privacy concerns. Advancements in the DyASR system have the potential to substantially enhance individuals' ability to effectively convey their intended communication, thereby improving overall comprehension.

In the literature, researchers have explored various strategies to enhance the performance of DyASR systems. To address the limited availability of atypical speech data, some studies, like \cite{artificial_data}, have investigated the use of artificial data generated through non-linear speech tempo modification, demonstrating performance improvements. Additionally, domain adaptation of pre-trained ASR models, as discussed in \cite{adaptation}, has shown promise in mitigating the scarcity of sufficient atypical data. In another study, presented in \cite{enhance}, it was demonstrated that improving speech quality through denoising techniques applied to atypical speech can transform disordered speech into a form more akin to normal speech. The study in \cite{bonet} compared traditional ASR with Neural Network-based ASR using features like Filterbanks (FBANKs) and Mel Frequency Cepstral Coefficients (MFCCs) for dysarthric speech. An investigation into raw magnitude spectra-based multi-stream acoustic modeling, which exhibited notable performance gains for dysarthric speech, was documented in \cite{raw_mag}. In a recent study, researchers explored the combination of raw phase-based representations with raw magnitude spectra, as described in \cite{raw_phase}, employing single and multi-stream architectures composed of cascades of convolutional, recurrent, and fully-connected layers for acoustic modeling. Furthermore, a Deep Neural Network (DNN) model utilizing MFCC-based i-vectors was found to outperform other models, including convolutional neural networks, gated recurrent units, and long short-term memory networks, as reported in \cite{dnn_ivector}. In \cite{decode_limit}, Hermann and Doss demonstrated that the Lattice Free - Maximum Mutual Information (LF-MMI) setup of F-TDNN surpassed the Time Delay Neural Network with Long Short Term Memory (TDNN-LSTM) model in recognizing dysarthric speech when using MFCC features.

To the best of our knowledge, no studies have been conducted to determine the selection of acoustic features and acoustic modeling techniques for dysarthric speech. In this study, we propose a systematic investigation of DyASR systems using spectral features with various acoustic models. 
The main contributions of our study includes: \vspace{-0.1cm}
\begin{itemize}
 \setlength{\itemsep}{-0.5pt}
\item A systematic investigation of spectral features and their combinations for acoustic models in the DyASR study.  
\item An assessment of the incorporation of pitch features in different systems.
\item An evaluation of the impact of changing the number of overlapping frames in training the F-TDNN acoustic model.
\item The introduction of individualized speaker performance assessment and severity labeling for optimal configuration.
\end{itemize}
\section{EXPERIMENTAL SETUP}
\label{sec:expsetup}
\begin{figure}[htb]
\centering
\includegraphics[height=3.5cm,width=7.5cm]{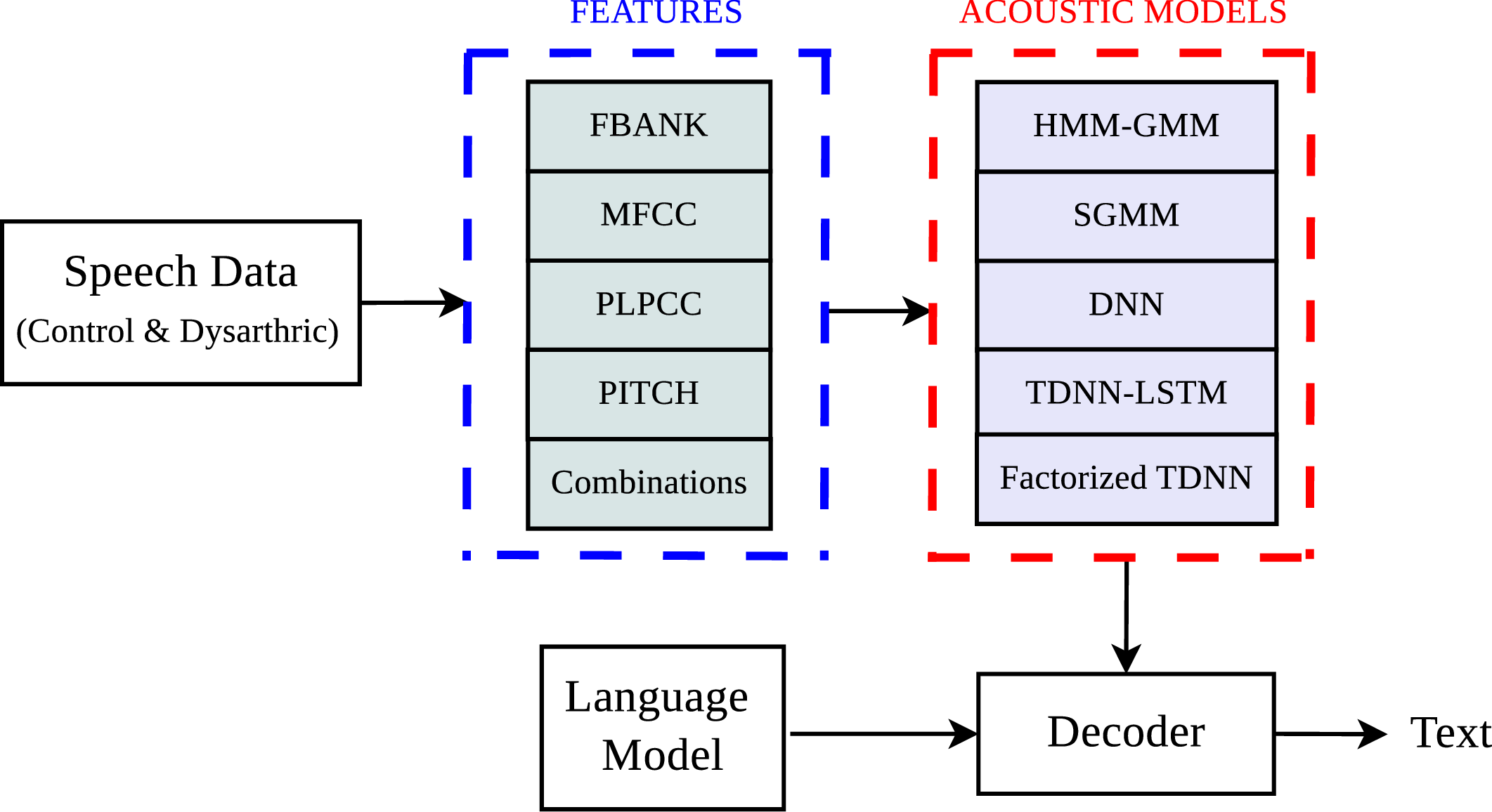}
\vspace{-0.1cm}
\caption{Experimental setup with various spectral features and acoustic models investigated in this study.}
\label{fig:flow}
\end{figure}

Figure \ref{fig:flow} depicts our experimental setup, which includes various features and acoustic models under investigation. We employed four types of acoustic features: FBANKs, MFCCs, Perceptual Linear Prediction Cepstral Coefficients (PLPCCs), as well as combinations with Pitch features. These features were evaluated using five different acoustic models: Hidden Markov Model with Gaussian Mixture Model (HMM-GMM), Sub-space Gaussian Mixture Model (SGMM), Time Delay Neural Network with Long Short Term Memory layers (TDNN-LSTM), and Factorized-TDNN (F-TDNN).

All experiments were conducted using the Kaldi speech processing toolkit \cite{kaldi}, with the recipe available in \cite{decode_limit}. It's worth noting that increasing the frame shift duration for dysarthric speakers during feature extraction has demonstrated improved performance \cite{bonet}. In this study, we maintained the same frame shift durations, i.e., 15 ms for dysarthric speakers and 10ms for control speakers.
\vspace{-0.4cm}
\subsection{TORGO Dysarthric Speech Database}
\label{ssec:data}
We are utilizing the publicly available TORGO database \cite{torgo}, a crucial resource for dysarthric speech research. The freely accessible TORGO database consists of a total of 15 hours of speech, comprising 8 speakers with dysarthria contributing approximately 6 hours of speech data and 7 control speakers contributing approximately 9 hours of speech data. In total, there are 16,394 utterances, and the statistical distribution of unique utterances showed that out of 971 unique utterances, 615 utterances are single-worded and 356 utterances are sentenced.

\vspace{-0.1cm}
\subsection{Selection of features}
\label{ssec:feature}
For preliminary investigation, we selected three spectral features \cite{fr1,fr2}. They are: FBANKs \cite{fbank}, MFCCs \cite{mfcc}, Perceptual Linear Prediction Cepstral Coefficients (PLPCCs) \cite{plpcc}.
Prior research on dysarthric speech has revealed noticeable and consistent differences in pitch levels between dysarthric speech and typical speech, particularly during spontaneous conversations \cite{pitch}. 
Consequently, we conducted an investigation into the impact of appending pitch features \cite{kaldi_pitch} to each of the three features, resulting in an additional three sets of features.

Initially, we trained the F-TDNN system using FBANKs, MFCCs, and PLPCCs features independently, without combining them. We adhered to standard practices, selecting 40-dimensional high-resolution features for both FBANKs and MFCCs, as per the F-TDNN model's requirements. When training with PLPCCs, we retained the original 13-dimensional configuration.

Furthermore, in our quest for an optimal feature combination, we opted for 40 high-resolution FBANKs features to complement the 13-dimensional features of MFCCs and PLPCCs with 3-dimensional Pitch features. This choice was made deliberately to strike a balance, ensuring that our feature set wouldn't become excessively large. This consideration is especially crucial when working with models trained on limited datasets, as overly complex features can pose challenges.
\vspace{-0.6cm}
\subsection{Acoustic model configurations}
\label{ssec:acoustic}
We investigated five distinct acoustic models to address the intricacies of DyASR. First, the HMM-GMM model utilized triphone modeling with context information and Gaussian Mixture Models comprising 400 components. This approach was complemented by forced alignment during training and subsequent decoding. Moving forward, the SGMM model further improved modeling capabilities through Subspace Gaussian Mixture Models \cite{sgmm}, featuring 8000 leaves and 19000 sub-states. The DNN model \cite{kaldi} was configured with five hidden layers, employing mini-batch training with 5000 mix-ups and parameter tuning across 20 epochs. The Time-Delay Neural Network with Long Short-Term Memory (TDNN-LSTM) \cite{kaldi} model introduced temporal dependencies with LSTM layers and employed chunk-based training to enhance performance.

The F-TDNN model was designed with a structure resembling that of TDNN-LSTM, with a particular focus on the integration of online i-Vectors \cite{ftdnn}. The corpus size is relatively limited, indicating a small-scale dataset. Furthermore, noticeable variations in speech tempo have been observed, primarily within the subset of dysarthric speakers. To address this, we applied a speed perturbation data augmentation technique \cite{speed_perturbation}. This involved modifying temporal rates to 0.9 and 1.1 relative to the native speaking rate. Additionally, we employed variable frames-per-chunk during training.

The training and evaluation followed a leave-one-out approach, where fourteen speakers were included in the training set, and one speaker was reserved for evaluation. Two distinct language models (LMs) were employed based on the specific task requirements. For the isolated word recognition task, a unigram (1-gram) LM was utilized, while a bigram (2-gram) LM was applied for the sentence recognition task. The decoding grammar was constrained to generate single-word outputs for the 1-gram LM. This constraint has been previously shown to be effective in completely mitigating insertion errors in prior research efforts \cite{decode_limit}. In the case of the F-TDNN model, we conducted experiments varying the degree of overlapping frames between training example chunks.
\vspace{-0.1cm}
\subsection{Overlapping frames between consecutive chunks}
\label{ssec:overlap}
Training a chain-based neural network acoustic model in Kaldi requires a focus on controlling contextual information \cite{chain}. We conducted experiments involving the manipulation of overlapping frames, ranging from 0 to 40 frames, to influence context consideration during the training process. This parameter adjustment has notable implications for memory consumption and the model's capacity to capture various aspects of speech variability, such as pronunciation, speaking rate, and acoustic characteristics. The incorporation of overlapping frames is crucial for mitigating speech variability within the chain-based model. In this model, each training instance consists of a chunk of frames containing extracted acoustic attributes.


\vspace{-0.2cm}
\section{RESULTS and Discussion}
\label{sec:result}

\begin{figure}[htb]
\begin{minipage}[b]{1.0\linewidth}
  \centering
\centerline{\includegraphics[height=4cm,width=8.5cm]{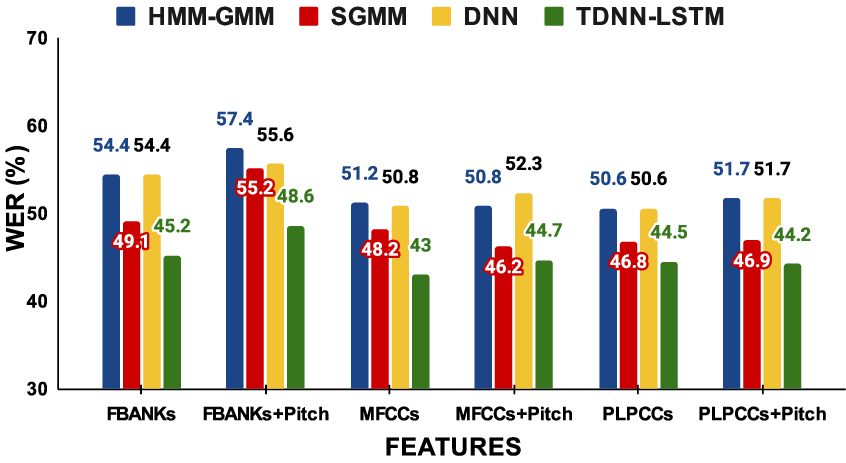}}
\end{minipage}
\vspace{-0.9cm}
\caption{Isolated word recognition performances  (in Word Error Rates (WERs)) for different acoustic models on dysarthric speech with various feature combination sets.}
\label{fig:iso}
\end{figure}

As discussed in Section \ref{ssec:acoustic}, for the dysarthric speech recognition task, we have divided it into isolated word and sentence recognition tasks based on their distinct applicability, aligning with previous research efforts \cite{decode_limit}. In the dysarthric isolated word recognition task, Figure \ref{fig:iso} illustrates the experimental results pertaining to three spectral features: FBANKs, MFCCs, and PLPCCs, along with their combinations with pitch features. These results were evaluated using the first four acoustic models (AMs) discussed in Section \ref{ssec:acoustic}, which include GMM-HMM, SGMM, DNN and TDNN-LSTM.

Figure \ref{fig:iso} provides valuable insights into the performance of different feature sets with various acoustic models.  For the HMM-GMM acoustic model, PLPCCs features delivered the best performance with a Word Error Rate (WER) of 50.6\%, closely followed by MFCCs features at 51.2\%. Interestingly, the addition of Pitch features did not yield significant improvements for PLPCCs; however, for MFCCs showed a slight 0.7\% relative improvement. The trend continued with the SGMM model, where PLPCCs features outperformed others with a WER of 46.8\% without pitch features. Again, the combination of Pitch features followed a similar pattern, resulting in a noteworthy 4\% reduction in WER for MFCCs, making it the top-performing feature for the SGMM model at 46.2\%. Moving to the DNN model, PLPCCs features led the way with a WER of 50.6\%, closely trailed by MFCCs with a 50.8\% WER. Meanwhile, for the TDNN-LSTM model, MFCCs emerged as the clear winner with a WER of 43\%. Importantly, Pitch features did not show significant improvements in the performance of neural network-based DNN and TDNN-LSTM AMs. Overall, the choice of features displayed distinctive trends across various acoustic models: HMM-GMM favored PLPCCs, SGMM performed best with MFCCs (with Pitch), DNN had a close competition between PLPCCs and MFCCs, and MFCCs excelled for the TDNN-LSTM model.



\vspace{-0.3cm}
\vspace{-0.6cm}
\begin{figure}[htb]
\begin{minipage}[b]{1.0\linewidth}
  \centering
\centerline{\includegraphics[height=4.5cm,width=8.5cm]{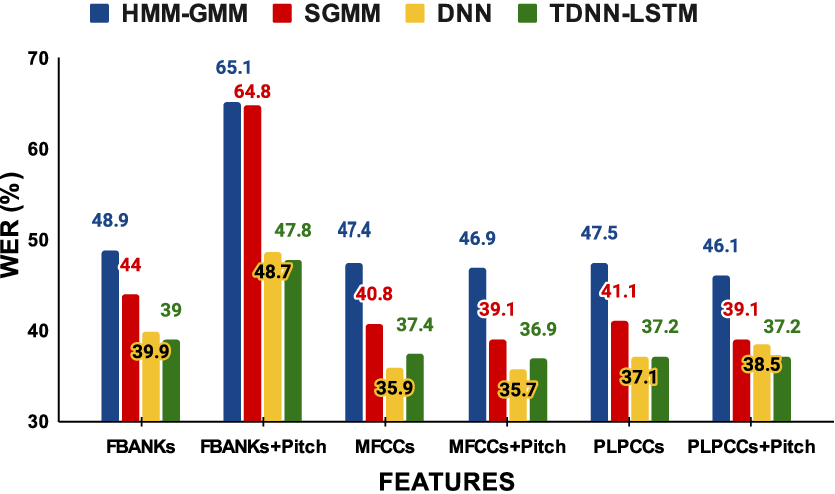}}
\end{minipage}
\vspace{-0.85cm}
\caption{Sentence recognition performances  (in Word Error Rates (WERs)) for different acoustic models on dysarthric speech with various feature combination sets.}
\label{fig:sent}
\end{figure}
\vspace{-0.3cm}
Figure \ref{fig:sent} provides a comprehensive view of the results for sentence recognition task. The HMM-GMM model excelled with MFCCs features, achieving a WER of 47.4\%. Notably, PLPCCs also performed impressively with a competitive score of 47.5\%, when evaluated without Pitch features. However, the introduction of Pitch features resulted in a notable 2.9\% improvement in the performance of PLPCCs, making it the top-performing feature set with a WER of 46.1\%. When Pitch features were not combined, the SGMM model delivered its best performance with MFCCs features, achieving a WER of 40.8\%, closely followed by PLPCCs with a WER of 41.1\%. Intriguingly, with the inclusion of Pitch features, the performance of PLPCCs showed a remarkable relative improvement of 4.9\%, matching the performance of MFCCs features (39.1\%) with Pitch. Moving to the DNN model, it demonstrated exceptional performance with MFCCs features, whether Pitch was included (WER of 35.7\%) or not (WER of 35.9\%). This result highlighted the robustness of MFCCs in this context. Interestingly, for the TDNN-LSTM model, the introduction of Pitch features did not significantly impact performance when used alongside PLPCCs. However, a slight relative improvement of 1.3\% was observed when Pitch features were added to the MFCCs features, resulting in the best WER of 36.9\% for the TDNN-LSTM model. Again in overall, the choice of features displayed distinct patterns across various acoustic models: HMM-GMM excelled with PLPCCs+Pitch, SGMM favored MFCCs+Pitch and PLPCCs+Pitch, DNN performed exceptionally with MFCCs+Pitch or without Pitch, and TDNN-LSTM showcased the best performance with MFCCs+Pitch.


In the case of the F-TDNN acoustic model, we conducted a comprehensive analysis encompassing both dysarthric (dys) and control (ctl) speakers, addressing both isolated word and sentence recognition tasks. Our exploration into frame overlap commenced with Kaldi's F-TDNN recipe, which, by default, assigns a null value for overlapping frames. In this investigation, we selected MFCCs as the baseline feature set. The baseline Word Error Rates (WERs) are conveniently summarized in Table \ref{tab:baseline}.

\vspace{-0.5cm}
 \renewcommand{\arraystretch}{1.2}
\begin{table}[htb]
\centering
\caption{Baseline WERs of F-TDNN Acoustic model with default value of 0 overlapping frames, on isolated word and sentence recognition tasks.}
\vspace{-0.3cm}
\label{tab:baseline}
\scalebox{0.9}{
\begin{tabular}{|c|c|c|c|c|} 
\hline 
\bf{Feature} & \bf{Isol(dys)} & \bf{Sent(dys)} & \bf{Isol(ctl)} & \bf{Sent(ctl)} \\
\hline
MFCCs & 54.4 & 38.5 & 43.2 & 13.0 \\ 
\hline 
\end{tabular}
}
\end{table}
\vspace{-0.3cm}

\begin{figure}[htbp]
\begin{minipage}[b]{1.0\linewidth}
  \centering
\centerline{\includegraphics[height=4.1cm,width=8.5cm]{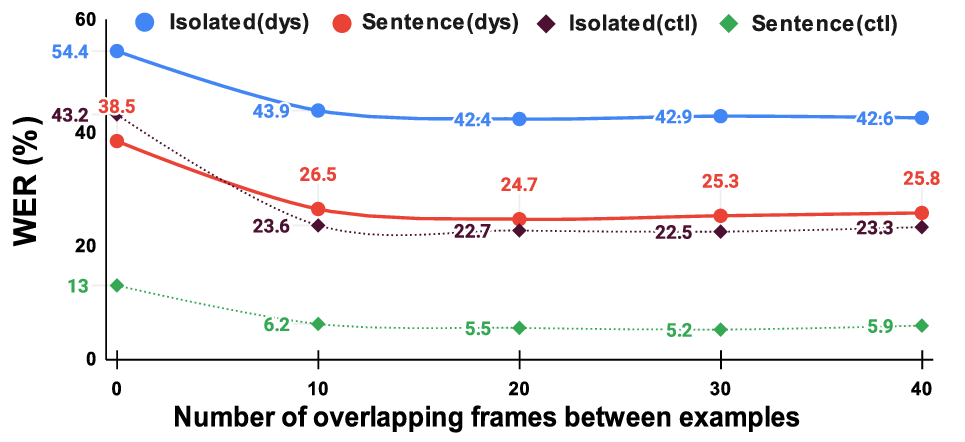}}
\end{minipage}
\vspace{-0.8cm}
\caption{Effect of variation in overlapping frames on F-TDNN setup with MFCCs features.}
\label{fig:frame_overlap}
\end{figure}

Subsequently, we extended our experimentation with the F-TDNN model, systematically varying the frame overlap from 0 to 40 frames, with a step size of 10 frames. The insights derived from Figure \ref{fig:frame_overlap} reveal that 20 frames of overlap yielded the optimal performance for the dysarthric speech recognition task, whether it involved isolated words or sentences. Furthermore, we also explored the effects of frame overlap variations on control speech. Remarkably, as illustrated in Figure \ref{fig:frame_overlap}, a 30-frame overlap demonstrated superior performance compared to other values in the case of control speech. However, it is essential to emphasize that our primary focus remains on dysarthric speech recognition. As a result, we selected the configuration with 20 frames of overlap between example chunks for further experimentation within the F-TDNN architecture.

 \vspace{-0.5cm}
\vspace{-0.3cm}
We thoroughly examined various feature combinations for the F-TDNN acoustic model, and the results are presented in Table \ref{tab:comparison}. Analyzing the findings from Table \ref{tab:comparison} for the dysarthric speech recognition task, it becomes evident that the most effective feature combination for isolated word recognition was achieved by combining FBANKs with MFCCs and Pitch features. This combination exhibited a remarkable 4.7\% relative improvement compared to \cite{decode_limit}, achieving a Word Error Rate (WER) of 41\%. Similarly, for the sentence recognition task, MFCCs emerged as the optimal feature set, showcasing a noteworthy improvement with a WER of 24.7\%, representing a relative gain of 4.6\%. Overall, our feature selection strategy for the F-TDNN model consisted of FBANKs+MFCCs+Pitch for isolated word recognition and MFCCs alone for sentence recognition.
\vspace{-0.7cm}
 \renewcommand{\arraystretch}{1.3}
\begin{table}[htbp]
\centering
\caption{Comparison of WERs using the frame overlap of 20 in the F-TDNN acoustic model with the LF-MMI setup \cite{decode_limit}. This evaluation is conducted for various combinations of acoustic features, for both Dysarthric and Control speech datasets.}
\vspace{-0.3cm}
\label{tab:comparison}
\scalebox{0.68}{
\begin{tabular}{|c|c|c|c|c|}
\hline
\bf{Features} & \bf{Isol
(DYS)} & \bf{Sent(DYS)} & \bf{Isol
(CTL)} & \bf{Sent(CTL)} \\
\hline
 MFCCs (LF-MMI setup) \cite{decode_limit} & 43.0 & 25.9 & 22.0 & 7.9 \\ \hline \hline 
FBANKs & 43.7 & 27.3 & 24.1 & 6.7\\
\hline
MFCCs & 42.4 & \textbf{24.7} & 22.7 & 5.5\\
\hline
PLPCCs & 42.7 & 24.8 & 22.0 & \bf{3.7}\\
\hline
FBANKs+Pitch & 43.3	& 27.9 & \bf{17.9}& \bf{3.7}\\
\hline
MFCCs+Pitch & 44.2 & 26.3 & 24.8 & 7.0\\
\hline
PLPCCs+Pitch & 42.4 & 25.8 & 22.0 & 5.6\\
\hline
FBANKs+MFCCs+Pitch & \textbf{41.0} & 26.0 & 21.7 & 5.2\\
\hline
FBANKs+PLPCCs+Pitch & 41.8 & 26.8 & 23.8 & 7.9\\
\hline
MFCCs+PLPCCs+Pitch & 44.3 & 27.6 & 24.0 & 7.5\\
\hline
FBANKs+MFCCs+PLPCCs+Pitch & 41.9 & 25.5 & 23.2 & 6.0\\
\hline
\end{tabular}%
}
\end{table}
\vspace{-0.3cm}

It is noteworthy that our approach, featuring 20 frames of overlap, has yielded superior results compared to the previous study \cite{decode_limit} across various feature sets. Notably, when we combined FBANKs, MFCCs, and Pitch features, we observed a substantial 4.7\% relative improvement in isolated word recognition tasks. Furthermore, in the context of sentence recognition tasks, the inclusion of MFCCs led to a remarkable 4.6\% relative improvement.

While our primary focus centered on dysarthric speakers, we also extended our evaluation to control speakers. The results, as presented in Table \ref{tab:comparison}, indicate that when the F-TDNN system was tested on control speakers, the WER demonstrated a noteworthy 18.6\% relative reduction, particularly in the case of isolated word recognition, where the combination of FBANKs and Pitch features excelled. For the recognition of sentence utterances among control speakers, once again, the combination of FBANKs with Pitch features delivered the best performance, showcasing a significant 53.2\% relative improvement. These outcomes underscore the adaptability of our approach, allowing us to select suitable feature and model combinations based on specific task requirements for further enhancing our study.

 \vspace{-0.6cm}
 \renewcommand{\arraystretch}{1.5}
 \begin{table}[htb]
\centering
\caption{ Speaker-specific performance analysis conducted on dysarthric speakers using F-TDNN Acoustic model for two tasks: Isolated word and Sentence recognition. The first and second rows displays speaker-wise WERs with the best features for isolated word recognition and sentence recognition tasks.}
\vspace{-0.3cm}
\label{tab:speaker-specific}
\scalebox{0.66}{
\begin{tabular}{|c|c|c|c|c|c|c|c|c|c|}
\hline
\bf{Features} & \bf{Task} & \bf{F01} & \bf{F03} & \bf{F04} & \bf{M01} & \bf{M02} & \bf{M03} & \bf{M04} & \bf{M05} \\
\hline 
\multirowcell{2}{FBANKs+MFCCs\\+Pitch} & \bf{Isol} & 55.32 & 39.72 & 12.85 & 50.62 & 53.95 & 9.02 & 59.68 & 46.68\\
 & \bf{Sent} & 50.54 & 10.69 & 2.08 & 30.05 & 26.79 & 3.70 & 38.68 & 31.67\\
 \hline 
\multirow{2}{*}{MFCCs} & \bf{Isol} & 65.96 & 38.85 & 14.06 & 52.41 & 57.73 & 8.03 & 57.88 & 43.99 \\
 & \bf{Sent} & 47.55 & 7.60 & 1.25 & 28.42 & 31.30 & 2.63 & 35.99 & 25.82 \\
\hline
\end{tabular}
}\vspace{-0.5cm}
\end{table}

\vspace{0.25cm}
In Table \ref{tab:speaker-specific}, we present speaker-specific WERs for the two best-performing feature sets in the F-TDNN model, which achieved the highest scores in isolated and sentence recognition tasks, respectively. Among the speakers, those denoted as F04 and M03 exhibited the lowest WERs, warranting the classification of 'mildly severe.' For speakers F03 and M05, their WERs fell within the range of 30-50\%, leading to their categorization as 'moderately severe.' Speakers with WERs exceeding 50\% have been classified as 'highly severe.' It's important to note that these severity labels are derived from the WERs obtained in the isolated word recognition task, providing valuable insights into the individual performance characteristics.
\vspace{-0.5cm}
\section{CONCLUSIONS}
\label{sec:typestyle}
In this study, we investigated various acoustic features and their combinations with four different acoustic models (HMM-GMM, SGMM, DNN, and TDNN-LSTM) for DyASR, encompassing both isolated words and sentences. Among these models, TDNN-LSTM achieved the lowest WER of 44.2\% in the isolated word recognition task, while DNN outperformed the others with a WER of 35.7\% in sentence recognition. Our study underscores the importance of selecting suitable acoustic features tailored to specific acoustic models. Notably, pitch features played a vital role in enhancing sentence recognition, particularly for dysarthric speech.

We also optimized a chain-based F-TDNN model by adjusting the number of overlapping frames between consecutive training chunks, resulting in improved performance. Our focus on dysarthric speech recognition led us to choose an optimal 20-frame overlap between training examples, which outperformed previous research \cite{decode_limit}. For isolated word recognition, the combination of FBANKs with MFCCs and Pitch achieved a WER of 41\%, representing a relative improvement of 4.63\%. In sentence recognition, MFCCs delivered the best performance with a WER of 24.7\%, showcasing a relative improvement of 4.65\%. Importantly, these improvements extended to control speech datasets.

Future work aims to further enhance DyASR by addressing speech variations, employing speech enhancement methods, and applying augmentation techniques to compensate for the limited dataset size. The preliminary investigation was conducted using a Wav2Vec2-based end-to-end ASR system, where various data augmentation techniques were explored for fine-tuning \cite{wav2vec2_Dys_Asilomar}. 


\vspace{-0.2cm}
\bibliographystyle{IEEEbib}
\bibliography{strings,refs}

\end{document}